# A new eye segmentation method based on improved U2Net in TCM eye diagnosis

Hong Peng[a], Niu Ning[b], Yilin Zhang[c], Guanjun Wang[*], Chenyang Xue[*]

[a]State Key Laboratory of Marine Resource Utilization in South China Sea, Hainan University, Haikou City, CHINA 570100; [b]State Key Laboratory of Marine Resource Utilization in South China Sea, Hainan University, Haikou City, CHINA 570100; [c]Department of Biomedical Engineering, Hainan University, Haikou City, CHINA 570100;

[*] Corresponding author: wangguanjun@hainu.edu.cn

[*] Corresponding author: xuechenyang@nuc.edu.cn

## ABSTRACT

Image segmentation is a very important technology in many fields such as image processing, pattern recognition, and artificial intelligence, and it is also the first step and an important key step in computer vision technology. For the diagnosis of Chinese medicine, tongue segmentation has reached a fairly mature point, but it has little application in the eye diagnosis of Chinese medicine.First, this time we propose Res-UNet based on the architecture of the U2Net network, and use the Data Enhancement Toolkit based on small datasets, in which the Res-U2Net network (based on PyTorch) relies on a large amount of data augmentation to use the available annotation samples more efficiently, and replaces the skip-connection module with residual soft on the basis of the original U2Net The connection module, which, in addition to reducing the semantic difference between low-level and high-level features, convolutional feature extraction can also be trained to reduce noise passed from low-level features to high-level features. Finally, the feature blocks after noise reduction are fused with the high-level features.Finally, the number of network parameters and inference time are used as evaluation indicators to evaluate the model. At the same time, different eye data segmentation frames were compared using Miou, Precision, Recall, F1-Score and FLOPS. To convince people, we cite the UBIVIS. V1[1] public dataset this time, in which Miou reaches 97.8%, S-measure reaches 97.7%, F1-Score reaches 99.09% and for 320×320 RGB input images, the total parameter volume is 167.83 MB,Due to the excessive number of parameters, we experimented with a small-scale U2Net combined with a Res module with a parameter volume of 4.63 MB, which is similar to U2Net in related indicators, which verifies the effectiveness of our structure.which achieves the best segmentation effect in all the comparison networks and lays a foundation for the application of subsequent visual apparatus recognition symptoms.

## 1. INTRODUCTION

In traditional Chinese medicine theory, medical visual diagnosis belongs to the category of visual diagnosis, which refers to the observation of different traits of scleral blood in the patient's eyes to diagnose certain diseases. The sclera of a normal eye is white because it belongs to the lungs and the lungs are white from the perspective of traditional Chinese medicine. However, if you have a certain disease, it may change to other colors, for example,Li[2] found a specific white eye signal of colorectal cancer, with a top bruising point or a mist-spotted shadow circle near the iris end at 5 or 7 points of the sclera, and a black stasis in the middle, the diagnostic compliance rate reached 76.9\%. The Zhu discovery can be based on the abnormal signals on the eye, such as eyelid Xanthoma, which can determine the new and old severity of the disease in diabetic patients.

In the mention of Yao medical visual diagnosis[4], and this diagnostic method is based on the visual diagnosis of traditional Chinese medicine combined with the characteristics of the Traditional ethnic medicine to further extend the visual diagnosis method, the content of which includes the white eye diagnosis method, the black eye diagnosis method, and the eye meridian area diagnosis method. White eye diagnosis is to observe the changes in the color, morphology and other changes of blood vessels on the sclera and bulb conjunctiva to determine the location, etiology, pathology, and prognosis of the disease. If the blood vessels in the corresponding area are angry, most of them are blood stasis or the disease is more severe and urgent, such as acute pneumonia, acute hepatitis, etc.; If you see that there is a lot of blood in the eye, and the color is light, most of them are Asthenia and Chills; If blue-black spots or blue-black spots appear in any part of the white eye, it is ascariasis; At the upper end and edge of the capillary between the sclera and the bulbous conjunctiva, a light purple and cloudy patch in an irregular form is hookworm disease. As a non-invasive, intuitive, and rapid diagnosis, visual diagnosis has been widely used in TCM diagnostics. Visual diagnosis has been employed extensively in TCM diagnostics since it is a non-invasive, simple, and quick method of diagnosis.

With the advancement of artificial intelligence technology in medical treatment[5][6], the intelligent processing algorithm that uses computer vision to diagnose pathological structures has achieved certain results. However, the four diagnoses of traditional Chinese medicine's tongue diagnosis are the basis for the majority of current research. For this reason, the purpose of this research is to build a single device based on standardized environmental conditions to collect the eyes at a fixed distance and put the collected eye images into the intelligent algorithm for training. Because we are in our own dedicated visual instrument. Therefore, the conditions can be moderately relaxed compared to the length of reasoning and the total number of model parameters, and the focus is on the partition accuracy.

In a fixed environment[7], we can get to contain more valid information and less noise, while in the traditional threshold segmentation, template matching[8][9] and area segmentation[10][11] can not be compared to the neural network processing effect on the image. In the field of deep learning, U-net[12] does not require thousands of annotated training samples compared to Deeplab-v3[13]. The U-net network structure algorithm can rely on the powerful use of data augmentation to use the available annotation samples more efficiently, while the U-net structure consists of a shrink path for extracting the context and a symmetrical extended path that can achieve precise positioning, so that high-level features can be extracted, but low-level feature information can be retained, but in recent years there is a new simple and powerful deep network architecture, U2-net, for significance object detection (SOD), The U2-Net system is a two-tier nested U-shaped structure. This structure has the following advantages:

(1) Since the residual U-block (RSU) we propose is mixed with receptor domains of various sizes, allowing for the acquisition of more contextual information at various sizes.

(2) The depth of the the entire architecture is increased without noticeably raising computational costs thanks to the pooling procedures performed in these RSU blocks. Using this architecture, we can train deep networks without using the backbone of an image classification problem. Since the majority of SOD networks follow a similar design pattern. Utilize existing backbone networks, such as Alexnet, VGG, Resnet, and Densenet, to extract deep features; however, because these networks were initially created for image classification, they primarily extract features that represent semantic meanings rather than local details and global contrast information, both of which are essential for significance detection. As a result, U2-Net is akin to creating a new network for SOD, allowing for training from scratch, and achieving comparable or better performance than networks based on existing pre-trained skeletons.

(3)Because it is used for scleral vascular segmentation in the future, the accuracy of the segmentation of the early orbit will be more stringent. In this study, we found in the first comparison that in the existing image segmentation network, the Miou accuracy of U2Net has reached the highest, but whether it is U-Net or for U2Net, skip connection is used to connect the encoding layer and the decoding layer, and we replace the skip-connection module of U2Net with the residual connection module. To reduce noise passed directly to the decoding module via skip connection. In terms of multi-resolution feature fusion, a multi-resolution salient image is obtained from the bottom to the top layer through significant image fusion, and the effective information of each layer is retained in the last layer of fusion. In addition, considering that the amount of data collected by the subsequent use of our own Visual diagnostic instruments is not large enough, it is easy to cause overfitting. Therefore, we build a new loss function based on focal loss through the proportion of the eye in each image. And weight the target segmentation area of the image with lower quality of the eye, to increase the loss value and strengthen the attention to the target area.

(4)However, U2Net network has an inevitable problem that the parameter volume is too large, resulting in too slow inference time. We introduce Res-Module combined with the simplified version of U2Net to fully realize the accuracy similar to the large volume of U2Net and ensure that the parameter volume is low, suitable for medical computer-aided diagnosis real-time requirements.

The remainder of this article consists of six parts.Related work is discussed in brief in Section 2.Section 3 describes the various types of semantic segmentation models. Section 4 describes the details of the experiment and reports the results. The experimental results are discussed in Section 5. Finally, the conclusion is drawn in Section 6.

## 2. RELATED WORK

### 2.1 Traditional morphological methods

Derakshani et al..[14] proposed a novel ocular biometric model based on the conjunctival vasculature. In their method, scleral cutting is done manually, and a multiscale regional growth method is used to identify the vasculature. In 2007, Derakhshani.[15]proposed to represent and match vascular structures using ocular surface texture design based on wavelet analysis and neural networks. In 2012, Zhou[16] submitted an automatic segmentation method based on pixel threshold.In 2011, the authors[17] propose scleral cutting using the K-means clustering method.

### 2.2 Deep learning

In 2018 and 2019, Rot[18] combined deep learning techniques with SBVPI public datasets to achieve good segmentation results, showing the results obtained by U-Net, RefineNet and its proposed model SegNet, and comparing them with other classical techniques[19]. In 2019, Wang[20] submitted a ScleraSegNet work based on the U-Net architecture to experiment with the UBIRIS.V2 and MICHE datasets.In the field of TCM vision for the use of visual diagnosis, we are on the inspiration of tongue diagnosis and then segment the eye image, we compared three classic mainstream models, U-Net++, U2Net, DeeplabV3+ in this experiment, of which U2Net is currently the most novel and very good network in the existing network. We improved the skip-connection module on the basis of the U2Net network and achieved better Miou accuracy than U2Net.

## 3. METHODOLOGY

The network structure is shown in Figure 1. This is the upstream task network of the traditional Chinese medicine eye diagnosis system, based on the U2net image segmentation network. And the downstream network will use the segmented images to identify eye symptoms in the follow-up network. And this time our work is mainly to design the upstream network to carry out Eye image segmentation, this directly determines the quality of the subsequent downstream network diagnosis effect.

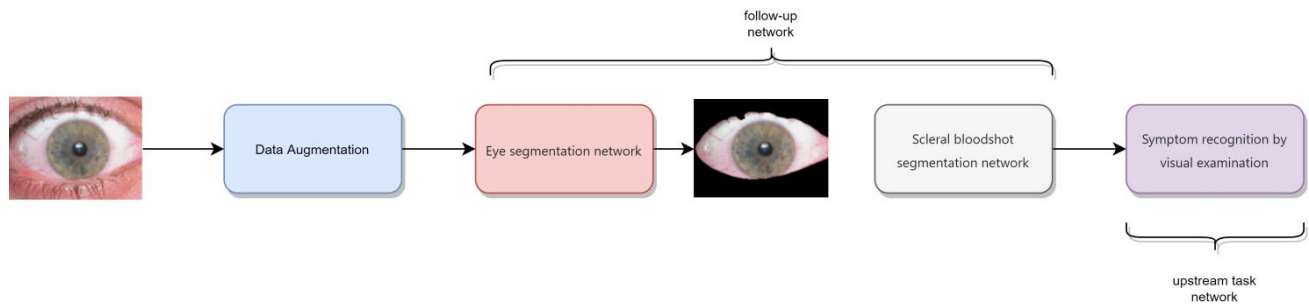

Figure 1: Res-U2Net-Lite-based eye image segmentation flowchart plus Downstream task recognition network flowchart.

### 3.1 Segmentation Flowchart

The segmentation flowchart of human eye face image based on Res-U2Net-Lite model. The system can provide medical image cutting based on the pupil, iris, and sclera of the human eye. This lays a certain foundation for the subsequent extraction of eyeball blood and then the task of TCM eye diagnosis and recognition in the downstream network.The network structure of u2net is shown in the eye segmentation network in Figure 1.

### 3.2 CNN

The 2012 Image Recognition Network Recognition Competition brought the academic study of CNNs to a pinnacle with the progression of these years. CNNs have so far created a system architecture that is more comprehensive. Convolutional layers, pooling layers, nonlinear layers, and fully linked layers are all added to the unique structure of CNN. Additionally, a lot of later networks, including those used for instance segmentation, semantic segmentation, and picture segmentation, are built on the connection and modification of the aforementioned layers.

### 3.2.1 Deeplabv3+

Chen publishes Deeplabv3+,[21] improves pyramid-shaped hole pooling, cascades multiple hole Convolution, and makes extensive use of batch normalization. Compared to previous networks, Deeplabv3+ has the following characteristics: 1. To continue to use the ASSP structure, SPPNET uses multi-ratio values to mine multiscale contextual content information with different resolution characteristics of multi-effective sensing fields. The decoding structure adds a new decoding module to reconstruct the boundary information after gradually reconstructing the spatial information to better capture the object boundaries. To decrease the number of parameters, try utilizing the enhanced modulev2 module as the network's core. Figure. 2shows the DeepLab V3+ network design. In Figure. 2, ASSP and $1\times1$ convolution and 4 times up sampling, and then the features obtained from the backbone network are matched with $1\times1$ convolution for the number of channels, and then the deep features are stitched with the shallow features obtained from backbone for feature fusion, which draws on the idea of residual block.

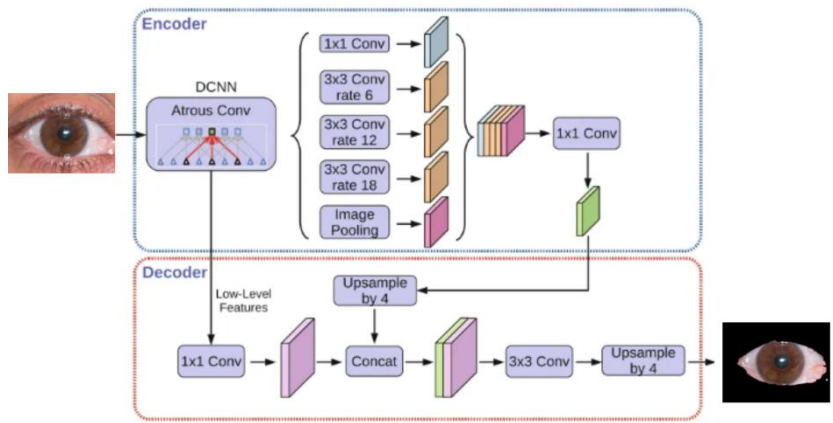

Figure 2: The network of Deeplabv3+ model.

### 3.2.2 UNet++

Unet++ was proposed by Zhu[22], Encoder and Dcoder are the two main components of UNet++, and they are coupled by a number of nested dense convolutional blocks. Prior to fusion, the major goal of UNet++ is to close the semantic gap that exists between the feature maps of the encoder and decoder. An example of this is the use of a dense convolution block with three convolution layers to fill the semantic gap between $(X^{0,0}, X^{1,3})$. In the graphical abstract, the original U-Net is shown in black, dense convolution blocks on the skip paths are shown in green and blue, and deep supervision is shown in red. UNet++ can be distinguished from U-Net by its red, green, and blue components. Figure 3 . depicts the Unet++ network structure.

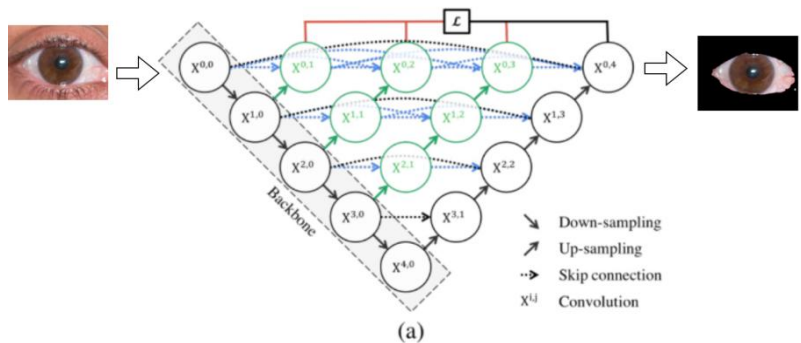

Figure 3: The network of UNet++ model.

### 3.2.3 U2Net

U-shaped networks for different tasks have been explored for some time,[23] and these methods are collectively referred to as "(U×n-net)" in the subsequent order to stack similar U-Net modules to build cascading models, where n is the number of duplicate U-net modules. While U2-Net networks are computationally and memory are also amplified n times. Theoretically, the exponent n can be set to any positive integer to achieve a single-level or multi-level nested U-shaped structure. But schemas with too many nesting levels would be too complex to implement and use in real-world applications. And u2-net sets n to 2, as shown in Figure 4 two-layer nested U structure, its top layer is a large U-shaped structure composed of 11 stages, each stage is composed of a wellconFigured residual block. Therefore, nested U structure can better extract multiscale features at all levels and scale fusion inter-level features.As Figure 4 shows, U2-Net is mainly composed of three parts: six encoders, five decoders significant Figure fusion feature modules. In the encoder phase, we use RSU (Residual U-block) RSU-7, RSU-6, RSU-5, and RSU-4. As mentioned earlier, L in RSU-L is the height of the RSU block, and L is usually conFigureured according to the spatial resolution of the feature map entered. For feature maps with larger heights and widths, we use larger L's to capture larger scale information. And the RSU block is shown as Figure 7. The cascade structure of the decoder is similar to the corresponding coding layer, and the input of each decoder comes from the previous level of the Upsample feature map and the cascade of the Upsample feature map from the symmetric encoder stage as input. The last layer is the significance feature fusion module, which is used to generate a significant plot probability map, we first receive six side significant outputs from six decoding layers plus RSU4-F through the $3 \times 3$ convolution and Relu function, and then map the respective corresponding significant feature maps into the sigmoid function, and then use $1 \times 1$ convolution to change the mapped feature map to the same latitude, and then perform the final stitching. And the RSU-4F block is shown as Figure 8.

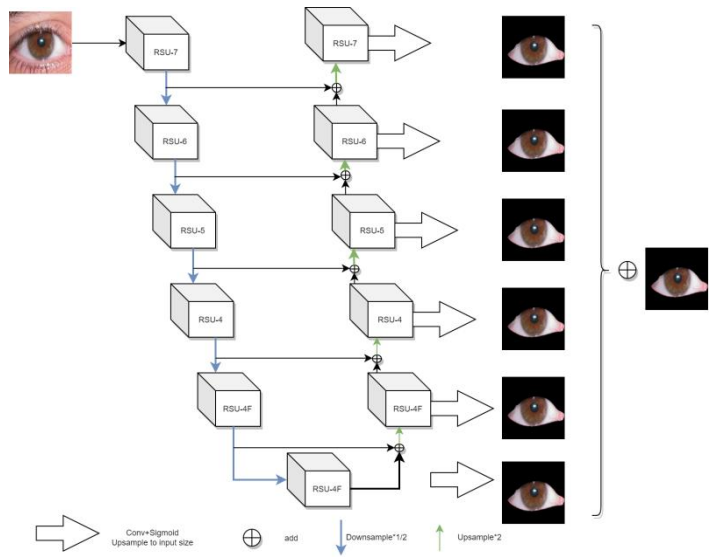

Figure 4: The network of U2Net model.

### 3.2.4 Res-U2Net

In U2Net, the RSU module of its sibling encoding layer and the RSU module of the decoding layer are connected via skip-connection. An RSU module is essentially a U-Net network, and U-Net itself has already performed multiple convolution and pooling operations on the input image. At this time, the output is high-level semantic features, losing a lot of spatial information. So we connect a residual connection block before entering an encoded RSU module to bridge the output of the RSU module with the input of the original RSU module.[25]And if there is noise in the input image in an RSU module, the noise will be transmitted by feature fusion at each layer. Inspired by the use of an improved residual connection in the RSU module of the encoding layer and decoding layer on the original U2Net network, as shown as Figure 5. Unlike the remaining paths, the residual soft-link module does not use overly complex convolutional layers. The whole module starts with the feature fusion part, which fuses the initial feature map of each stage with the coded feature map. The coding layer is an RSU module, which Max-Pooling the output of the RSU module and the original input of the RSU module and then closely follows the 1 × 1 convolutional block. And the connection coefficient is flexible, which can greatly increase the nonlinear features and enhance the information fusion between channels while keeping the scale of the feature map unchanged. Thereby increasing the semantic features of the incoming feature map. At the same time, the parameter change of the convolution kernel can effectively control the information transfer of the remaining soft connections. Training convolution kernels can reduce invalid feature transfers. In addition to reducing the semantic differences between low-level and high-level features, convolutional feature extraction can also be trained to reduce noise from low-level features to high-level features. Finally, the denoted feature blocks are fused with the high-level features.

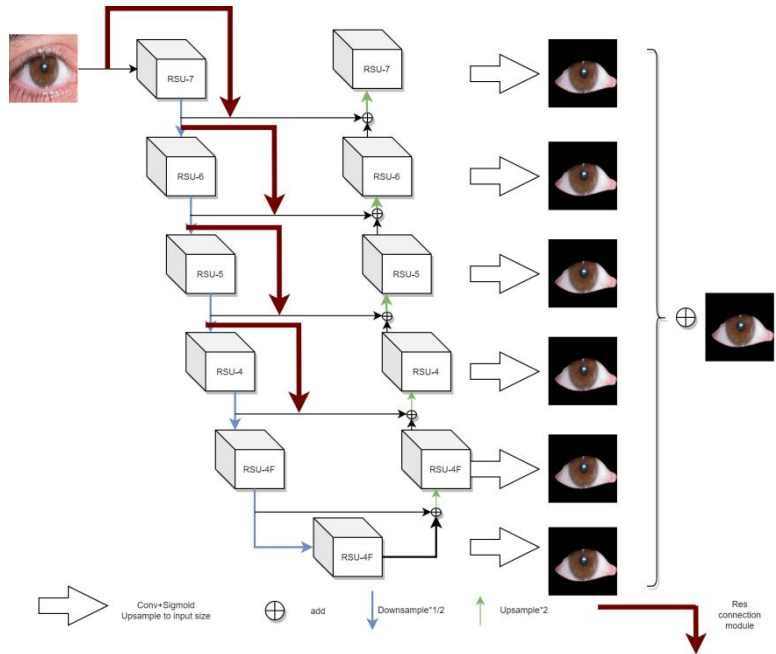

Figure 5: The network of Res-U2Net model.

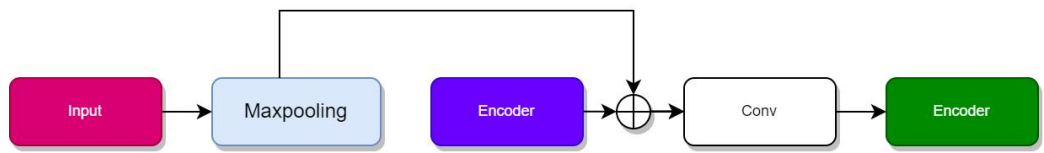

Figure 6: The network structure of Res connection module.

## 3.3 Database and experimental environment

### 3.3.1 Data Augmentation
We also leveraged the Data Enhancement Toolkit to efficiently scale up the sample size and make the trained network more generalizable, which greatly improved the robustness of the model. We used the Transform in Torchvision to mirror flip, flip up and down, zoom in, and rotate the data.

### 3.3.2 Dataset
We used the public dataset of UBIRIS.V1[1],which consists of 1877 images collected from 241 people in 2004 at two different sessions. The shooting conditions are: the camera model is Nikon E5700, and we selected the RGB image with 800×600 resolution as experimental data. Among them, 800×600 upper eye face images a total of 1205.because we extract the eye sclera, iris after extracting the eye sclera, iris after the segmentation of the picture. And then we should segment the scleral blood vessels.so we choose the higher resolution of the eye face image for the network to train the segmentation for the subsequent effect of the better. We decided 80% for the training dataset and 20% for the test dataset.See in Table 1.

Table 1: The number of specific test set and verification set

| Classifification | Quantity |
|---|---|
| Training Dataset | 964 |
| Test Dataset | 241 |

### 3.3.3 Experimental Environment

Our experiments were conducted on a high-performance computing platform. The machine is configured as follows: AMD R9-6900HX 3070TI 8GB ideo memory; The system environment is Linux, Python, PyTorch. The learning rate of all networks is set to 0.001 to start, the model parameters are updated using the AdamW gradient descent method, and the warm-up descent method is sampled on the model to degenerate the learning rate.

### 3.3.4 Evaluation indicators

We used Recall, the Precision, and F1-score, as well as Miou, the most important metric for image segmentation, and we also used Flops, which measures model complexity, and the parameter indicator. In Tab. 2 we can clearly see the Confusion matrix According to the above four indicatorsthe calculation methods of sensitivity and specificity mentioned can be further derived, such as Eqs. (1) and Eqs. (2).

$$Recall = \frac{TP}{TP+FN} \qquad (1)$$

Table 2: Confusion matrix

| Actual value | Positive (P) | Negative (N) | Total |
|---|---|---|---|
| Positive (T) | True positive (TP) | True negative (TN) | TP + TN |
| Negative (F) | False positive (FP) | False negative (FN) | FP + FN |
| Total | TP + FP | TN + FN | TP + TN + FP + FN |

$$\text{Precision} = \frac{TP}{TP+FP} \qquad (2)$$

$$\text{Miou} = \frac{1}{k+1}\sum_{i=0}^{k}\frac{TP}{FP+FN+TP} \qquad (3)$$

$$\text{F1-Score} = \frac{2*Recall*Precision}{Precision+Recall} \qquad (4)$$

$$\text{MAE} = \frac{1}{H*W}\sum_{r=1}^{H}\sum_{c=1}^{W}|P(r,c)-G(r,c)| \qquad (5)$$

In this experiment, this indicator is used to determine the fitting effect of the mask image segmented by the model and the eye, and the higher the accuracy, the better the fitting effect. In this experiment, we take the values of Ground truth and the predicted image as inputs to the calculation Miou. such as Eqs. (3).The F1 score is the sum of the averages of the two weighted weights. And the higher the F1 score, the more robust the model becomes.such sa Eqs. (4).MAE is the mean absolute error and represents the mean pixel difference between the predicted significance plot and its Ground Truth,such sa Eqs. (5).

# 4. RESULTS

## 4.1 Segmented Image Results

Figure 9 shows the results of Deeplabv3 + , UNet++ , U2Net , Res-U2Net and Res-U2Net-Lite of the eye.We can clearly see that U2net's segmentation effect on the image is closest to the artificially labeled image, and in Figure 9, Res-U2net has the best effect on the upper eyelash segmentation and the smoothness of the segmentation in all networks.And in our early experimental planning, the interference of eyelashes was removed when manual labeling. This is conducive to the division of blood vessels in the later stages, so it is normal that the segmentation of the upper eyelid will be slightly missing.

## 4.2 Performance comparison of the CNN models

Table 3 shows the specific performance metrics for Deeplabv3+ , UNet++ , Res-U2Net and Res-U2Net-Lite, and it can be seen that the Res-U2Net metric works best among all the comparison networks. Figure 10 shows the superiority of the three networks and intuitively shows that the Res-U2Net network achieves the optimal segmentation index of the eye compared to other networks.

## 4.3 Operational Performance Comparison

As can be seen in Table 4, in the same operating environment, The trained network is compared to the split image time, and the size of the model training memory is compared. As can be seen from the performance results, although Res-U2Net achieves the optimal segmentation index of the eye compared to other networks, However, in subsequent applications, our device is a fixed visual instrument. And the Res-U2Net-Lite's execution time has a gap compared with other models, And the accuracy of Res-UNet-Lite is not lower than that of U2Net. which has great forward-looking significance for the subsequent segmentation of scleral blood vessels and the diagnosis of corresponding diseases.

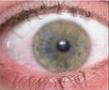

Figure 9: Actual segmentation results of each model.

Table 3: Comparison with algorithms based on CNNs.

| Model | Recall(%) | Precision(%) | Miou(%) | MAE | F1-Score(%) |
|---|---|---|---|---|---|
| UNet++ | 95.6 | 92 | 95.5 | 0.066 | 92.80 |
| Deeplabv3+ | 97.83 | 98.30 | 95.9 | 0.057 | 98.20 |
| U2Net | 99.05 | 98.48 | 97.30 | 0.0147 | 98.90 |
| Res-U2Net | 99.10 | 99.25 | 97.8 | 0.009 | 99.09 |
| Res-U2Net-Lite | 99.00 | 98.40 | 97.20 | 0.012 | 98.90 |

Table 4: Model performance

| Model | Inference Time(s/per image) | Parameters(MB) | Flops(G) |
|---|---|---|---|
| UNet++ | 0.083 | 34.960 | 54.537 |
| Deeplabv3+ | 0.078 | 5.813 | 20.651 |
| U2Net | 0.128 | 167.83 | 58.830 |
| Res-U2Net | 0.146 | 172.810 | 67.54 |
| Res-U2Net-Lite | 0.070 | 4.630 | 21.754 |

## 5. DISCUSSION

Image recognition and segmentation have always been hot issues in the field of deep learning. The core of image recognition research is the qualitative analysis of images, and the core of image segmentation lies in the qualitative analysis of pixel-level images.The images studied in this article should be used for visual identification in traditional Chinese medicine, which is different from the general images. In traditional Chinese medicine visual examination, the blood of the eye represents the location of each lung, so the accuracy of the early cutting is particularly important.After the orbit of the eye is divided, the subsequent blood of the eyeball is also divided again. Therefore, the image recognition and scleral region segmentation algorithm for the first extraction of the segmentation accuracy of the scleral region of the eye are particularly important. Based on the U2net network structure, this paper proposes a solution for the segmentation of eye area in the early stage for visual diagnosis in traditional Chinese medicine.As far as we know, today's work has a relatively good algorithm for iris segmentation.Lian[25] proposed a way to add the attention mechanism module to the U-net network for iris segmentation. Zhang[26] proposed an Unet network for interference to the Iris segmentation.However, the image required by TCM visual dialectics is not just an iris image. As mentioned earlier, our job is to cut the human eye image including the sclera iris together to facilitate subsequent disease recognition.In contrast to those SOD models built on Backbone,U2-Net is built on the proposed RSU-L Block, RSU-L Block can obtain local and global contextual information, which is important for segmentation tasks.And in Table 3 can be clearly observed that Res-U2Net's experimental effect has reached its best.But it doesn't take as long to perform as Res-UNet-Lite .The next step will revolve around the following aspects:
(1) Under the condition that the segmented area of the eye has been significantly improved, the blood crestless in the 14

areas of the eye and the characteristics of related visual diagnosis of traditional Chinese medicine are extracted.
(2) We will use our own TCM visual examination instrument to collect our own data set to segment, form our own data set and carry out subsequent related experiments.

## 6. CONCLUSION

As in traditional Chinese medicine, visual examination mainly relies on the direct contact of clinicians to conduct face-to-face examinations of patients. Time-consuming, laborious and prone to fatigue misdiagnosis, and neural network technology has directly injected new vitality into the medical industry. This paper proposes an upstream visual segmentation model based on U2Net neural network, and achieved good experimental results, our model's Recall reached 99.0%, Precision reached 98.40%, and the Miou index reached 97.2%, the shortest real-time segmentation in all networks, which laid the foundation for subsequent eyeball blood splitting. But there is still a long way to go to complete the design of the entire system. The model in this paper is currently only experimenting on publicly available datasets, and accurate segmentation is only the beginning of intelligent visual consultation. In our future research plan, we hope to use North University of China's Visual diagnostic instruments to collect our own eye data set and use this network architecture for experiments and related improvements. And then build the eyeball blood triage network after the segmentation effect reaches the required accuracy.

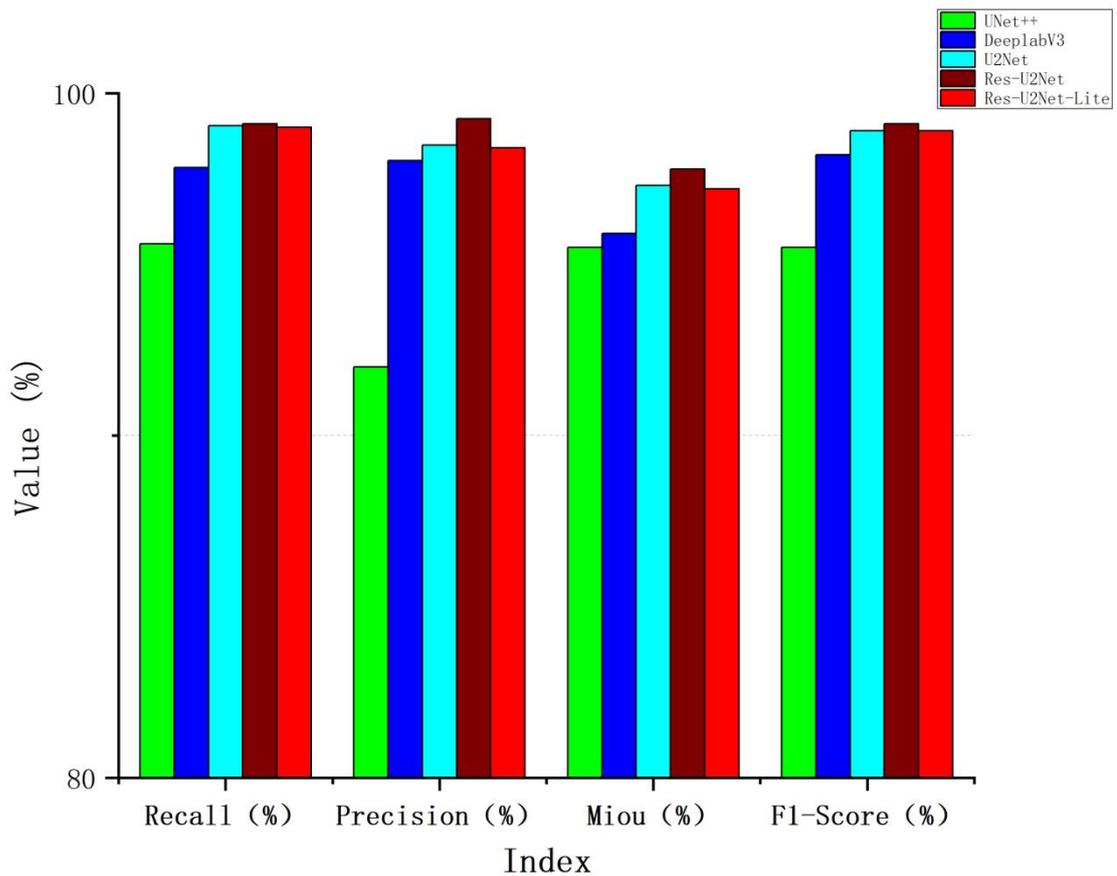

Figure 10: Performance comparison using Deeplab v3 +, UNet++, U2Net ,Res-U2Net ,Res-U2Net-Lite.

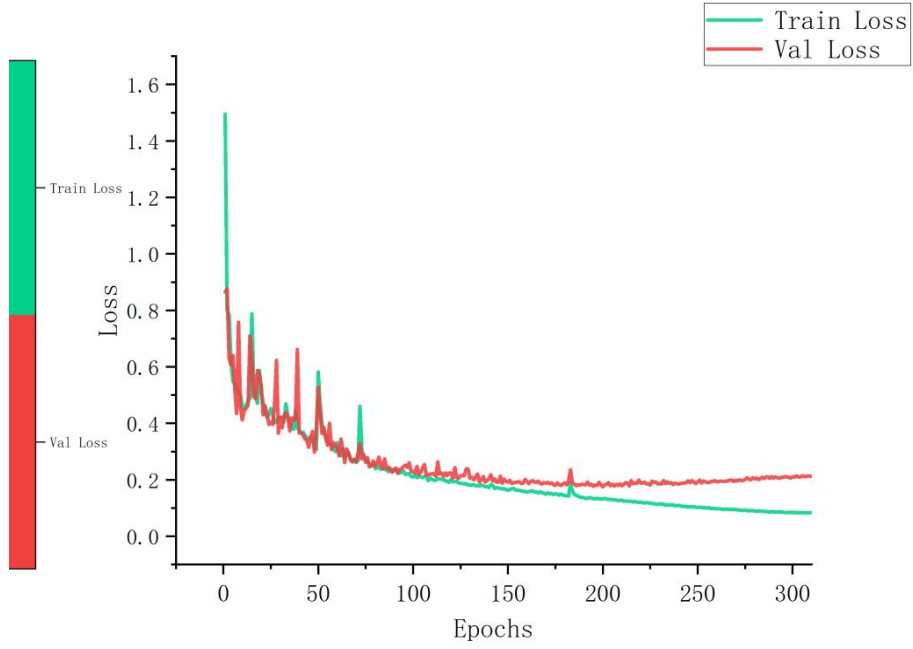

Figure 11: Loss in Res-U2net

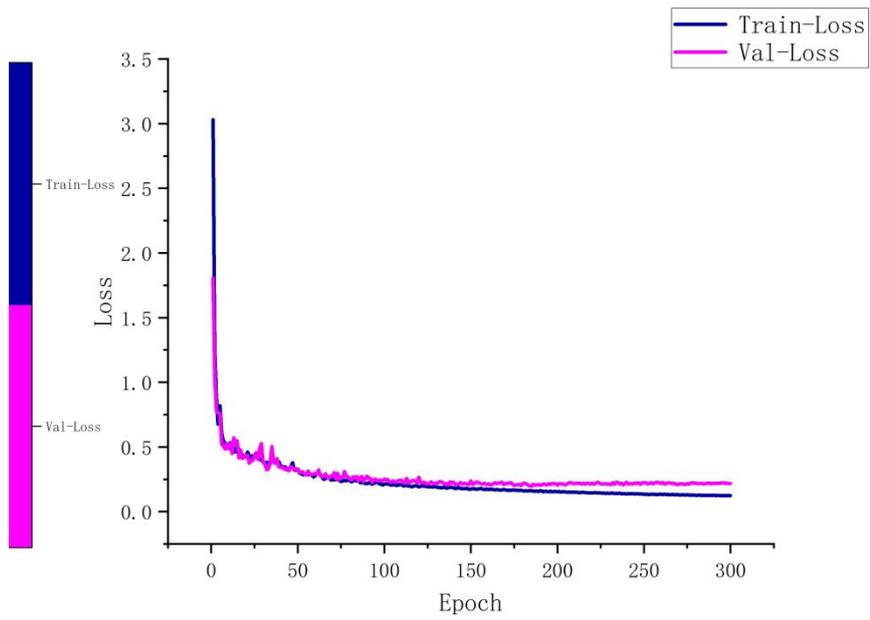

Figure 12: Loss in Res-U2net-Lite

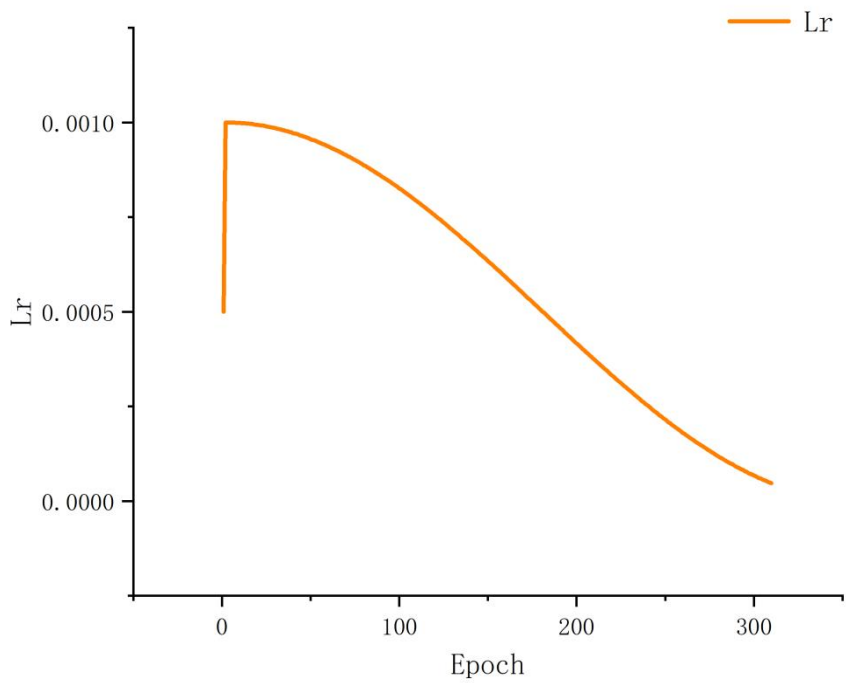

Figure 13: Learning Rate in Res-U2Net

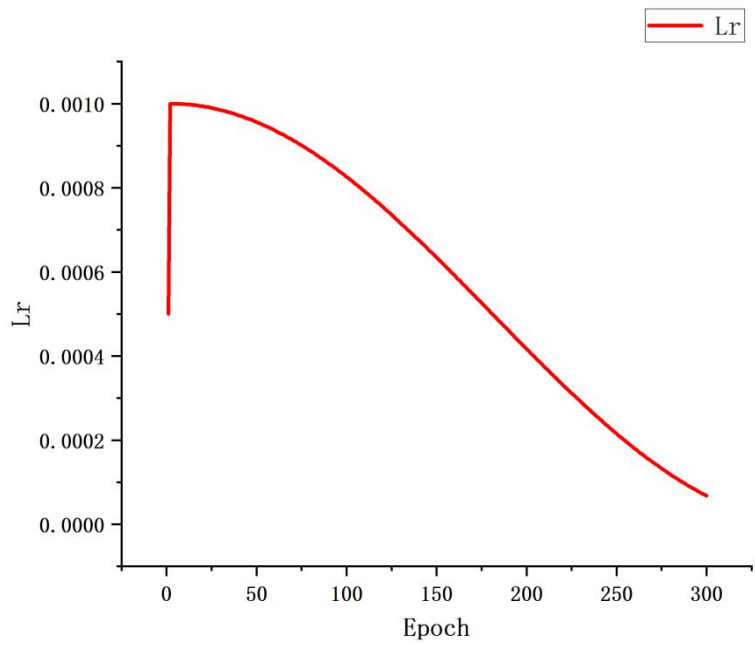

Figure 14: Learning Rate in Res-U2Net-Lite